\documentclass[aps,prd,superscriptaddress,nofootinbib,floatfix,twocolumn]{revtex4}
\usepackage{graphicx} 
\usepackage{amsmath}

\newcommand{\beq}{\begin{equation}}
\newcommand{\eeq}{\end{equation}}
\newcommand{\beqa}{\begin{eqnarray}}
\newcommand{\eeqa}{\end{eqnarray}}
\newcommand{\beqm}{\begin{multline}}
\newcommand{\eeqm}{\end{multline}}
\newcommand{\nnb}{\nonumber}

\begin{document}


\title{Exploring pseudo-Nambu--Goldstone bosons by stimulated photon colliders
in the mass range 0.1~eV to 10~keV}


%
\author{Kensuke Homma}
\affiliation{Graduate School of Science, Hiroshima University, Kagamiyama, Higashi-Hiroshima, Hiroshima 739-8526, Japan} 
\affiliation{International Center for Zetta-Exawatt Science and Technology, Ecole Polytechnique, Route de Saclay, F-91128 Palaiseau Cedex, France}
\author{Yuichi Toyota}
\affiliation{Graduate School of Science, Hiroshima University, Kagamiyama, Higashi-Hiroshima, Hiroshima 739-8526, Japan}
%


\date{\today}

\begin{abstract}
Searching for pseudo-Nambu--Goldstone bosons (pNGBs) in weak-coupling domains
is crucial for understanding the dark components in the universe.
We propose searching for pNGBs coupled to two photons 
in the mass range from 0.1~eV to 10~keV.
This could provide opportunities to test string-theory-based pNGBs 
beyond the GUT scale $M \sim 10^{16}$~GeV included in the weak coupling 
proportional to $M^{-1}$.
We provide formulae that are applicable to photon--photon scattering via a pNGB resonance
exchange with a stimulation process in an asymmetric head-on photon--photon 
collider by mixing three laser pulses in laboratory experiments.
We discuss the quantum electrodynamic effects on the pNGB exchange in the same mass--coupling domain as a background process from the standard model.
We find that a large unexplored mass--coupling domain is accessible
by combining existing laser facilities, including free-electron lasers.
\end{abstract}


\maketitle

\section{Introduction}
Spontaneous symmetry breaking could be  one of the most robust guiding principles
for general discussions of dark components in the universe.
Whenever a global symmetry is broken, a massless boson may appear as a Nambu--Goldstone boson (NGB).
In nature, however, an NGB emerges as a pseudo-NGB (pNGB) with a finite mass.
Even if a pNGB is close to being massless, its decay into lighter particles 
such as photons is kinematically allowed.
The neutral pion is such an example of a pNGB via chiral symmetry breaking.
The effective interaction Lagrangian is defined as
\beq\label{eq1}
-{\cal L} = gM^{-1} \frac{1}{4}F_{\mu\nu}F^{\mu\nu} \phi,
\eeq
where 
we assume a scalar-type field $\phi$ for simplicity, and
$M$ has the dimension of energy whereas $g$ is a dimensionless constant.

There are several theoretical models that predict low-mass pNGBs
such as axions~\cite{axion},
dilatons~\cite{dilaton}, and string-theory-based axion-like particles~\cite{strings}. However, pinning down the physical mass of a pNGB by means of such models is 
commonly non-trivial. Therefore, experiments are indispensable for investigating
the physical mass as widely as possible in the lower mass range.

We have previously advocated a novel method~\cite{PTP-DE} 
for stimulating 
$\gamma\gamma \rightarrow \phi \rightarrow \gamma\gamma$ scattering in
a quasi-parallel collision system (QPS) via an s-channel resonant pNGB exchange
by utilizing the coherent nature of laser fields.
Kinematically, this is analogous to four-wave mixing 
in matter~\cite{FWM} if the nonlinear atomic process is replaced with 
the pNGB exchange.
A QPS is intended for searching for pNGBs in the sub-eV mass range.
Indeed, we have performed two previous laboratory searches~\cite{PTEP-EXP00,PTEP-EXP01}
based on this approach.

Recently, an unidentified emission line, $\omega \sim 3.5$~keV,
has been reported in the photon energy spectra from a single galaxy and galaxy 
clusters~\cite{Bulbul,Boyarsky}, and the arguments are still actively 
ongoing~\cite{Arguments}. The possible interpretation of a pNGB decaying into
two photons has been discussed~\cite{Limits-7keV}. 
Indeed, string theories predict pNGBs to be
homogeneously distributed on a log scale in the mass range 
possibly up to $10^8$~eV~\cite{strings}.
This situation motivates us to try to extend the same method up to 10~keV in general.
Given the effective Lagrangian ${\cal L}$,
we evaluate the sensitivity to 
$\gamma\gamma \rightarrow \phi \rightarrow \gamma\gamma$ scattering
by means of stimulation using coherent X-ray sources.
However, we emphasize that the formulae we provide here are nevertheless applicable
to a wide mass range from 0.1~eV to 10~keV 
by combining different types of coherent and incoherent light sources.
This mass range is yet to be probed intensively, and thus
the proposed approach will open 
a window through which to explore string-theory-based pNGBs
beyond the GUT scale $M \sim 10^{16}$~GeV.

\section{Formulae for stimulated asymmetric photon--photon scattering via pNGBs}
\begin{figure}
\includegraphics[scale=0.25]{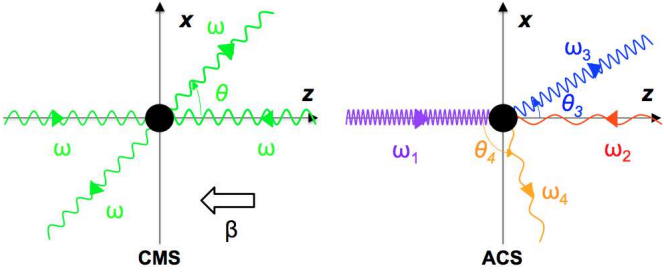}
\caption{\label{Fig1} 
Asymmetric collision system (ACS) realized in the laboratory frame, 
interpreted as the result of a Lorentz boost by a relative velocity $\beta$ of
the laboratory frame along the head-on collision ($z$) axis 
in the center-of-mass system (CMS). All four photons are confined within a
common reaction plane, the $x$--$z$ plane.} 
\end{figure}

To distinguish clearly between the photons in the colliding beams and the signal photons whose frequencies have been shifted by the scattering process,
we consider an asymmetric collision system (ACS)
as illustrated in Fig.~\ref{Fig1}. The ACS is obtained
by boosting a center-of-mass system (CMS) along the head-on
collision axis ($z$ axis) with a speed $\beta c$, where
$c$ is the speed of light 
(hence, the Lorentz factor is $\gamma = 1/\sqrt{1-\beta^2}$).
Four-momenta in the ACS are defined as
\beqa\label{eq_1}
p_1 = \omega_1(1, 0, 0, 1),
p_2 = \omega_2(1, 0, 0, -1), \mbox{\hspace{2.6cm}}\\ \nnb
p_3 = \omega_3(1, \sin\theta_3, 0,  \cos\theta_3),
p_4 = \omega_4(1, \sin\theta_4, 0, -\cos\theta_4).
\eeqa
For later convenience, with an incident photon energy 
$\omega$ in the CMS, we define the incident energies in the ACS as
$\omega_1 \equiv u\omega$ and 
$\omega_2 \equiv u^{-1}\omega$ via the following relationships:
$u \equiv \sqrt{(1+\beta)/(1-\beta)} = \gamma + \sqrt{\gamma^2-1}$.
With $u^+ \equiv u + u^{-1}$ and $u^- \equiv u - u^{-1}$, 
energy--momentum conservation results in the following relationships:
\beqa\label{eq_4}
&\mbox{0-axis: }& u^+\omega = \omega_3 + \omega_4,\\ \nnb
&\mbox{z-axis: }& u^-\omega = \omega_3\cos\theta_3 - \omega_4\cos\theta_4,\\ \nnb
&\mbox{x-axis: }& \omega_3\sin\theta_3=\omega_4\sin\theta_4.
\eeqa

We have formulated a QPS with a symmetric incident angle $\vartheta$,
which is the half angle between two incident photons
when a laser beam with the photon energy $\omega \sim$ 1~eV is 
focused by a lens element.
This formulation is also applicable to the CMS.
In the coplanar condition whereby
the plane determined by $p_1(1)$ and $p_2(1)$ coincides with that determined by
$p_3(1)$ and $p_4(1)$ 
(where the numbers in parentheses specify the linear polarization states), 
the Lorentz-invariant scattering amplitude of the scalar-field exchange is expressed as
\beq
{\cal M}_{S} =-(g M^{-1})^2\frac{\omega^4 \left(
\cos2\vartheta -1\right)^2}{2\omega^2 \left(
\cos2\vartheta -1\right)+m^2},
\label{mxelm_7}
\eeq
where the subscript $S$ denotes combinations of linear polarization 
states of four photons in the initial and final states.
The nonzero amplitudes are
${\cal M}_{1111}={\cal M}_{2222}=-{\cal M}_{1122}=- {\cal M}_{2211}$
where the linearly polarized states $(1)$ and $(2)$ are orthogonal to each other.
Since all the squared amplitudes are common, we omit the 
polarization subscript $S$ below.
In the following, the denominator of Eq.(\ref{mxelm_7}) is denoted by ${\cal D}$. We then introduce the imaginary part due to the resonance state
by the following replacement:
\beq
m^2 \rightarrow \left( m -i\Gamma/2 \right)^2 \approx
m^2 -im \Gamma,
\label{mxelm_9}
\eeq
where the decay rate $\Gamma$ is expressed as
\beq
\Gamma=(16\pi)^{-1} \left( g M^{-1}\right)^2 m^3
\label{eq_Gamma}
\eeq
for a given mass $m$~\cite{PTP-DE}.
Substituting this into the denominator in Eq.~(\ref{mxelm_7}) and
expanding around $m$, we obtain
\beq
\hspace{-.1em}{\cal D}\approx -2\left( 1-\cos2\vartheta  \right) \left( \chi+ia
\right)\quad\hspace{-.7em}\mbox{with}\quad\hspace{-.7em} \chi =\omega^2 -\omega_r^2,
\label{mxelm_10}
\eeq
where
\beq
\omega_r^2 =\frac{m^2/2}{1-\cos 2\vartheta },\quad
a=\frac{m \Gamma/2}{1-\cos 2\vartheta}.
\label{mxelm_12}
\eeq
Because the scattering amplitude is Lorentz invariant,
we have only to consider $\vartheta = \pi/2$ in the above formulation,
corresponding to the CMS, in order to apply it to the ACS
with the arbitrary parameter $u$ and $u^{-1}$ associated with a Lorentz
boost along the head-on collision axis in the CMS.
In the CMS, Eq.~(\ref{mxelm_12}) is thus simplified as
\beq\label{eq_CMSa}
\omega_r^2 =\left(\frac{m}{2}\right)^2,\quad
a=\frac{m}{4} \Gamma.
\eeq

From Eqs.~(\ref{eq_Gamma}) and (\ref{mxelm_12}),  $a$ is also
expressed as
\beq\label{eq_a}
a = \frac{\omega^2_r}{16\pi}\left(\frac{g m}{M}\right)^2,
\eeq
which shows explicitly the proportionality to $M^{-2}$.
The squared amplitude is then expressed as
\beq
|{\cal M}|^2 \approx  (4\pi)^2 \frac{a^2}{\chi^2+a^2}.
\label{mxelm_13}
\eeq
In the off-resonance case $\chi\gg a$, 
equivalent to Eq.~(\ref{mxelm_7}),
$|{\cal M}|^2$ is largely suppressed because of the factor $a^2 \propto M^{-4}$ for
large $M$. In contrast,
in the limit of $\omega \rightarrow \omega_r$,
$|{\cal M}|^2 \rightarrow (4\pi)^2$ is expected from Eq.~(\ref{mxelm_13}).
In principle, this is independent of how small $a \propto M^{-2}$ is.

Physically, however, it is difficult to satisfy 
$\chi = 0$ exactly in the case of
an extremely small $a$ because $\Gamma$ becomes so narrow. 
A CMS energy uncertainty is caused by momentum (angle) and
energy uncertainties originating from the spatial and temporal localizations 
of the laser fields by the focusing and shortening pulses, respectively. 
This uncertainty due to the wavy aspect of a photon is unavoidable, 
even at the level of a pair of photons in principle.
For a given uncertainty $\chi_{\pm} \equiv \pm \eta a$
with $\eta \gg 1$,
we still can expect an enhancement from
the averaged effect~\cite{PDG} over the uncertainty as follows:
\beqa\label{eq_M2ave}
\overline{|{\cal M}|^2}
=
\frac{1}{\chi_+ - \chi_-} \int_{\chi_-}^{\chi_+}|{\cal M}|^2 d\chi
\mbox{\hspace{2.4cm}}\\ \nnb
= \frac{(4\pi)^2}{2\eta a}2a\tan^{-1}(\eta) 
= (4\pi)^2 \eta^{-1} \tan^{-1}(\eta)
\\ \nnb
\approx (4\pi)^2 \eta^{-1}\frac{\pi}{2} = 8\pi^3\frac{a}{|\chi_{\pm}|},
\mbox{\hspace{2.3cm}}
\eeqa
where the approximation is applied to the situation $\eta \gg 1$.
By introducing a beam-energy uncertainty $\Delta\omega$
resulting in a bandwidth $\omega_{\pm} \equiv m/2 \pm \Delta\omega$
in the CMS with respect to the resonance condition $\omega=m/2$, the uncertainty
in $\chi$ in the CMS is expressed as
\beq
\chi_{\pm} = \omega^2_{\pm} - (m/2)^2 = \Delta\omega^2 \pm m\Delta\omega
\approx \pm m\Delta\omega,
\eeq
where $\Delta\omega^2 \ll m\Delta\omega$ is assumed in the above approximation.
Therefore, with Eqs.(\ref{eq_Gamma}) and (\ref{eq_CMSa}), 
the average of the squared scattering amplitude is approximated as 
\beq\label{eq_Ms3}
\overline{|{\cal M}|^2} \approx 8\pi^3 \frac{a}{m\Delta\omega}
= \frac{\pi^2}{8\Delta\omega}\left(\frac{g}{M}\right)^2 m^3.
\eeq
Compared to the off-resonant case $|{\cal M}|^2 \propto a^2$,
$\overline{|{\cal M}|^2} \propto a$ still has a gain of $a^{-1} \propto M^2$.
This is one of the two most promising features of our approach.
We note that the uncertainty of the CMS energy is related to
the beam-energy uncertainties $\delta\omega_1$ and $\delta\omega_2$
in the ACS (laboratory frame) as follows:
\beq\label{eq_Domega}
\Delta\omega = \sqrt{\omega_1\omega_2}
\sqrt{\left(\frac{\delta\omega_1}{\omega_1}\right)^2 +
\left(\frac{\delta\omega_2}{\omega_2}\right)^2}
\equiv \frac{m R}{\sqrt{2}},
\eeq
where in the last step we assume that an experiment tunes the beam energies
so that they satisfy $2\sqrt{\omega_1\omega_2} = m$ with
a common relative energy uncertainty $R \equiv \delta\omega_i/\omega_i$
for $i=1,2$.
By substituting Eq.~(\ref{eq_Domega}) into Eq.~(\ref{eq_Ms3}),
we eventually express the mass and coupling dependence of the squared 
scattering amplitude for a given relative-beam-energy uncertainty $R$ as
\beq\label{eq_Ms2}
\overline{|{\cal M}|^2} \approx
\frac{\sqrt{2}\pi^2}{8R}\left(\frac{g}{M}\right)^2 m^2.
\eeq

We then parametrize the differential cross section $d\sigma_{ngb}$~\cite{PTEP-EXP00}
for a pNGB exchange in the ACS (laboratory frame) as
\beqa\label{eqYsigma}
d\sigma_{ngb} &=&
\frac{1}{K(\vartheta) 2\omega_1 2\omega_2}
\overline{|{\cal M}|^2} dL_{ips},
\eeqa
where 
\beq\label{eqLIPS}
dL_{ips} = (2\pi)^4 \delta(p_3+p_4-p_1-p_2)
\frac{d^3p_3}{2\omega_3(2\pi)^3}\frac{d^3p_4}{2\omega_4(2\pi)^3}
\eeq
and 
the relative velocity $K$ is defined as~\cite{K-factor}
\beqa\label{eqK}
K(\vartheta) \equiv
\sqrt{(\vec{v_1}-\vec{v_2})^2 -
\frac{(\vec{v_1} \times \vec{v_2})^2}{c^2}}
= 2c\sin^2\vartheta = 2 
\eeqa
for $\vartheta=\pi/2$ and $\hbar=c=1$.
With $d^3p_3 = |\vec{p_3}|\omega_3 d\omega_3 d\Omega_3$,
we can express the differential cross section as a function of 
the solid angle $d\Omega_3$ of signal photon energy $\omega_3$ as follows:
\beq\label{eq_10}
\frac{d\sigma_{ngb}}{d\Omega_3} =
\left(\frac{1}{8\pi\omega}\right)^2
\left(\frac{\omega_3}{2\omega}\right)^2 \overline{|{\cal M}|^2}
\eeq
with
\beq\label{eq_omega3}
\omega_3 \equiv \frac{2\omega}{u^+ - u^- \cos\theta_3},
\eeq
which is a result of energy--momentum conservation.

\begin{figure}
\includegraphics[scale=0.30]{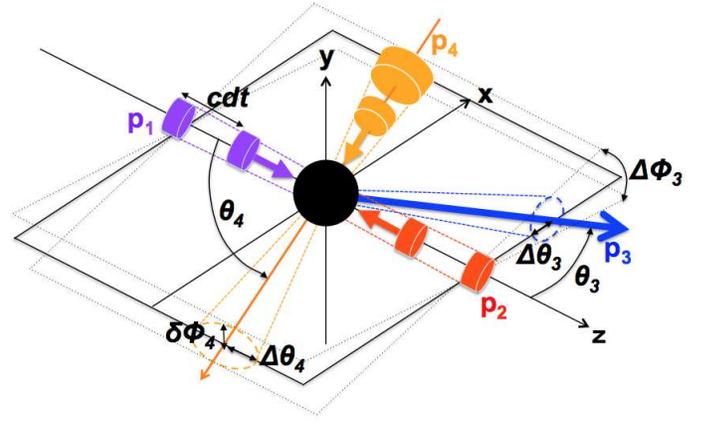}
\caption{\label{Fig2} 
Collision geometry between three pulsed photon beams.
Creation photon pulses $p_1$ and $p_2$ collide head-on at the origin
of the $xyz$-coordinates
and a coherent laser pulse $p_4$ to stimulate the scattering is simultaneously
focused into the origin. As a result of the stimulated 
$\gamma\gamma\rightarrow\gamma\gamma$ scattering, signal photons $p_3$ are
produced on the same reaction plane. The solid angle of $\vec{p}_3$ enhanced by the
inducing beam $p_4$ is related to the angular spread of $\vec{p}_4$ via 
energy-momentum conservation. This relationship
gives the rotation angle $\Delta\phi_3$ of the reaction plane around the $z$-axis.
} 
\end{figure}

In what follows, for simplicity, we assume perfect energy
resolution of signal photon $\omega_3$ so that we can discuss the
momentum range of $p_3$ based only on the angular spread of $\vec{p}_3$
from $\underline{\theta_3}$ to $\overline{\theta_3}$ on the reaction plane
and $\Delta\phi_3 \equiv \overline{\phi_3} - \underline{\phi_3}$
in the perpendicular direction with respect to the reaction plane 
in the ACS (see Fig.~\ref{Fig2}).
We now express the partially integrated cross section $\tilde{\sigma}_{ngb}$
over $d\Omega_3$ in the ACS as follows:
\beqa\label{eq_Sdm}
\tilde{\sigma}_{ngb} &=&
\frac{\overline{|{\cal M}|^2}}{(8\pi\omega)^2}
\int_{\underline{\phi_3}}^{\overline{\phi_3}} d\phi_3
\int_{\underline{\theta_3}}^{\overline{\theta_3}} 
\left(\frac{\omega_3}{2\omega}\right)^2 \sin\theta_3 d\theta_3
\mbox{\hspace{2cm}}
\\ \nnb
&=&\frac{\frac{\sqrt{2}\pi^2}{8R}\left(\frac{g}{M}\right)^2 m^2}{(8\pi\omega)^2}
\frac{\Delta\phi_3(\cos\underline{\theta_3}-\cos\overline{\theta_3})}
{(u^+ - u^- \cos\underline{\theta_3})(u^+ - u^- \cos\overline{\theta_3})}
\\ \nnb
&=& \frac{\sqrt{2}}{2(16\omega)^2 R}\left(\frac{gm}{M}\right)^2 {\cal I}
= \frac{\sqrt{2}}{128R}\left(\frac{g}{M}\right)^2 {\cal I},
\eeqa
where Eqs.~(\ref{eq_omega3}) and (\ref{eq_Ms2}) are substituted, and we introduce
\beqa\label{eq_I}
{\cal I} \equiv
\frac{\Delta\phi_3(\cos\underline{\theta_3}-\cos\overline{\theta_3})}
{(u^+ - u^- \cos\underline{\theta_3})(u^+ - u^- \cos\overline{\theta_3})}.
\eeqa
As shown in the last step of Eq.~(\ref{eq_Sdm}),
$\tilde{\sigma}_{ngb}$ eventually becomes 
independent of mass because the beam energies in the ACS are tuned 
so that $\omega=m/2$ is satisfied in the corresponding CMS.

The second important feature of our approach is to induce 
a specific two-photon final state by supplying an additional coherent
field with a different wavelength $\omega_4$ from any of $\omega_1$,
$\omega_2$, or $\omega_3$. 
Simultaneously, the momentum of the induction field $p_4$ must 
coincide with that of one of the two photons in the final state. 
Therefore, the stimulated range of the signal photon $p_3$ is determined 
by the momentum range of the inducing coherent photons $p_4$ after all. 
For simplicity, we assume that the momentum spread of $p_4$ is dominated 
by the angular spread of $\vec{p}_4$ due to the short focal length of a lens element
by neglecting the intrinsic energy spread of $\omega_4$.
In this case, we can express the angular range $\theta_3$ via the
third relation in Eq.~(\ref{eq_4}) as
\beqa\label{eq_}
\overline{\theta_3} = 
\sin^{-1}\left(\frac{v}{u^+ - v}\sin(\theta_4 + \Delta\theta_4)\right)
\\ \nnb
\underline{\theta_3} = 
\sin^{-1}\left(\frac{v}{u^+ - v}\sin(\theta_4 - \Delta\theta_4)\right),
\eeqa
where $\omega_4 \equiv v\omega$ and $\Delta\theta_4 \sim 1/(2F_4)$ 
by assuming that the $p_4$ beam is focused with the f-number $F_4$ 
as illustrated in Fig.~\ref{Fig2}. 
Figure~\ref{Fig2} depicts all the optical beam axes as being on
a common reaction plane, and $p_4$ is provided as a cone-like focused
beam based on Gaussian optics. Therefore, a slight deviation of $\vec{p}_3$ from the
coplanar condition is determined by the angular spread of $\vec{p}_4$.
This azimuthal angular spread, $\delta\phi_4$, of $\vec{p}_4$ from the reaction 
plane gives the rotation angles of the reaction planes around the $z$ axis.
Hence, we can express $\Delta\phi_3$ as
\beqa
\Delta\phi_3 =
2\sin^{-1} \left( \frac{\sin\delta\phi_4}
{\sqrt{\sin^2\theta_4\cos^2\delta\phi_4+\sin^2\delta\phi_4}} \right),
\eeqa
where $\delta\phi_4 = \Delta\theta_4$ holds because of the axial
symmetry around the optical axis of the $p_4$ beam. 
By substituting these parameters into Eq.~(\ref{eq_I}), 
we can evaluate an inducible partially integrated
cross section from Eq.~(\ref{eq_Sdm}).

With the parametrization for signal yield, ${\cal Y}$, 
developed in Appendix A of Ref.~\cite{PTEP-EXP00},
we express the expected number of signals with $\omega_3$ per pulse
crossing for the spontaneous scattering process as 
\beqa\label{eq_Ys}
{\cal Y}_s = 
\int dt \int dx^{i} \rho_1(t,x^{i}) \rho_2(t,x^{i}) K \left[1/L^2\right]
\tilde{\sigma}_{ngb} \left[L^2\right],
\eeqa
where $i$ runs from 1 to 3, 
and the first factor corresponds to time integrated beam luminosity
with the dimension of $1/L^2$ with length unit $L$.
We then express the induction effect by convoluting
the probability distribution function of an induction field $P_4$
as follows:
\beqa\label{eq_Yi}
{\cal Y}_{i} &=&
\int dt \int dx^{i} \rho_1(t,x^{i}) \rho_2(t,x^{i})
K \tilde{\sigma}_{ngb} P_4(t,x^{i}) 
\\ \nnb
&\equiv&
K \tilde{\sigma}_{ngb} N_1 N_2 N_4 G,
\eeqa
where 
$\rho_k$ indicates a normalized density distribution
over an infinite space-time range for creation beams $k=1,2$ and 
the induction beam $k=4$ by assuming pulsed Gaussian beams 
with the number of photons contained in individual pulses, 
$N_1$, $N_2$, and $N_4$, as defined explicitly in the following.
The factor $G$ is the integrated geometrical overlap factor.
At a space point $x^i$ for a given common time $t$ in the ACS,
the squared field strength of a Gaussian laser pulse can be parametrized 
as \cite{Yariv}
\beqa\label{eq_gauss}
I(x,y,z=ct) = E^2_0 \frac{w^2_0}{w^2(ct)}
e^{-2\frac{x^2+y^2}{w^2(ct)}}
e^{-2\left(\frac{z-ct}{c\tau}\right)^2},
\eeqa
where $E_0$ is the electric field and $E^2_0$ is proportional to
the number of photons $N$ in a pulse,
$\tau$ is the pulse duration time, and the beam radius 
$w(ct) = w_0\sqrt{1 + (ct/Z_R)^2}$ is 
given for wavelength $\lambda$,
beam waist $w_0=2F\lambda/\pi$, and 
Rayleigh length ${Z_R}=\pi w^2_0/\lambda$.
The volume for the normalization is thus obtained as
\beqa\label{eq_V}
V = \int^{\infty}_{-\infty} \frac{I(x^i)}{E^2_0} dx^i = (\pi/2)^{3/2} w^2_0 c \tau.
\eeqa
Based on the three-beam arrangement in the identical reaction plane
in Fig~.\ref{Fig2},
the normalized densities $\rho_k = I_k/V_k$ are defined as follows:
\beqa\label{eq_rhok}
\rho_1(x^{i}) =
N_1 \frac{(2/\pi)^{3/2}}{w^2_1(ct)c\tau_1}e^{-2\frac{x^2+y^2}{w^2_1(ct)}}
e^{-2\left(\frac{z-ct}{c\tau_1}\right)^2}, \mbox{\hspace{1.4cm}}
\\ \nnb
\rho_2(x^{i}) =
N_2 \frac{(2/\pi)^{3/2}}{w^2_2(ct)c\tau_2}e^{-2\frac{x^2+y^2}{w^2_2(ct)}}
e^{-2\left(\frac{z+ct}{c\tau_2}\right)^2}, \mbox{\hspace{1.4cm}}
\\ \nnb
P_4(x^{i}) = V_4 \rho_4(x^{i}) 
=
N_4\frac{{w^2_0}_4}{w^2_4(ct)}
e^{-2\frac{X^2+y^2}{w^2_4(ct)}}
e^{-2\left(\frac{Z-ct}{c\tau_4}\right)^2},
\eeqa
where $X \equiv \sin(\pi+\theta_4) x + \cos(\pi+\theta_4) z$ and
$Z \equiv \cos(\pi+\theta_4) x - \sin(\pi+\theta_4) z$ are introduced
for the rotated incidence of the coherent induction field with respect
to the head-on collision axis of the ACS.

The geometrical overlap factor integrated over the longest
pulse duration $\tau_L$ among the three pulses whose central
positions are all at the origin at $t=0$ is then expressed as
\beqa\label{eq_G}
G=(2/\pi)^{3/2}({w_0}_4/(c\tau))^2
\int^{\tau_L}_{-\tau_L} dt (w_1 w_2 w_4)^{-2} \times \\ \nnb
\{A(1/w^2_1+1/w^2_2+1/w^2_4)B\}^{-1/2} e^{-2Dt^2},
\eeqa
where 
for simplicity we assume that experiments introduce $\tau=\tau_1=\tau_2$ with
\begin{eqnarray*}\label{eq_ABD}
A \equiv 1/w^2_1 + 1/w^2_2 +
(\cos\theta_4/w_4)^2 + (\sin\theta_4/(c\tau_4))^2,
\mbox{\hspace{6cm}}
\\ \nnb
B \equiv (\sin\theta_4/w_4)^2 + (\cos\theta_4/(c\tau_4))^2
+ 2/(c\tau)^2 - \mbox{\hspace{6.6cm}} \\ \nnb
[\sin\theta_4\cos\theta_4\{1/w^2_4-1/(c\tau_4)^2\}]^2/A,
\mbox{\hspace{6cm}} \\ \nnb
D \equiv 2/\tau^2 + 1/\tau^2_4
-\left[ \sin^2\theta_4/A +
\mbox{\hspace{8.7cm}} \right. \\ \nnb
\cos^2\theta_4 \{1 + (1/w^2_4-1/(c\tau_4)^2)\sin^2\theta_4\}^2/B\
\left.\right] / (c\tau^2_4)^2. \mbox{\hspace{5cm}}
\nnb
\end{eqnarray*}
Because the total number $Y$ of detected signal photons is described as
$Y = f T \epsilon {\cal Y}_i$ with collision repetition rate $f$~[Hz],
data accumulation time $T$~[s], and detection efficiency $\epsilon$,
a reachable coupling strength is finally expressed as
\beqa\label{eq_goM}
\frac{g}{M} = 2^{1/4} 8 \sqrt{\frac{R Y}
{{\cal I}f T \epsilon K G N_1 N_2 N_4}}
\eeqa
from Eqs.~(\ref{eq_Sdm}) and (\ref{eq_Yi}).
We note that $g/M$ has $m^{-1}$ dependence 
because $G \propto \omega^2 \propto m^2$.

\section{Sensitivity and QED background}
\begin{figure}
\includegraphics[scale=0.45]{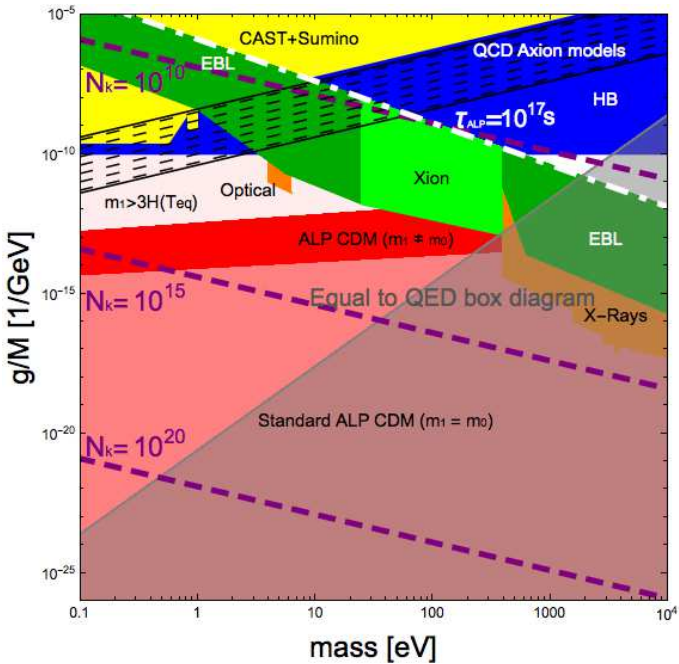}
\caption{\label{Fig3} Sensitivity to the mass--coupling domains.
Experimentally excluded and theoretically predicted domains are 
all imported from Ref.~\cite{WISPY}, where all details are explained.
A natural constraint that the lifetime of dark matter is equal to
the age of the universe at the shortest is expressed as the dash-dotted line.
The three dashed lines show accessible coupling limits
with $N_k=10^{10}, 10^{15}, 10^{20}$ common for $k=1,2,4$, respectively
based on Eq.~(\ref{eq_goM}) for the common parameter set in Table~\ref{Tab1}.
This figure shows the domain in which the QED process
dominates the cross section of a pNGB exchange observed by the same method.
The boundary line of the gray translucent area shows $m$--$g/M$, 
at which $\tilde{\sigma}_{ngb} = \tilde{\sigma}_{qed}$ is satisfied.
} 
\end{figure}

The three dashed lines in Fig.~\ref{Fig3} show $g/M$~[1/GeV] as a function of 
$m$~[eV] for $N_k=10^{10}, 10^{15}$, and $10^{20}$ for $k=1,2,4$, respectively
based on Eq.~(\ref{eq_goM}) in the range 0.1~eV to 10~keV.
The experimentally excluded and theoretically expected domains 
of the mass--coupling relationship are imported from Ref.~\cite{WISPY},
where all details are explained.
For this figure, the set of parameters summarized in Table~\ref{Tab1}
is assumed, where the parameters in the upper rows
can be commonly used for any $\omega = m/2$. 
We note here that a long f-number $F$ is assumed deliberately, thereby 
guaranteeing that the CMS energy does not fluctuate because of the uncertainty
in the incident angles but rather because of the energy spread
of the creation beams. This allows the use of nearly parallel incoherent
creation pulses, for instance, those generated by laser-electron Compton 
scattering, as we discussed in all-optical-based $\gamma$--$\gamma$ 
scattering~\cite{PTEP-GG}.
As an example of $\omega$, we show concrete energies and scattering 
angles in the second set of rows if we aim at $m=7$~keV 
in order to test the possibility of pNGB exchange
to explain the excess~\cite{Bulbul}.

Above the keV range, we must take photon--photon scattering based on
quantum electrodynamics (QED) into account 
as a background source from the standard model.
Probing QED-based photon--photon scattering via four-wave mixing
has been proposed in Ref.~\cite{FWMqed}. These calculations are based
on the Euler--Heisenberg effective Lagrangian and the signal yield is
evaluated within a classical wave picture. The formulation we provide here is
based on the particle picture of colliding photons and thus allows us 
to simply plug the partially integrated cross section, $\tilde{\sigma}_{qed}$,
for the QED box diagram into Eq.~(\ref{eq_Yi}), which has been applied to 
$\tilde{\sigma}_{ngb}$ on the same footing.
The differential cross section of the elastic QED scattering process~\cite{QED}
with respect to scattering angles in the ACS is expressed as
\beqa\label{eq_QED}
d\sigma_{qed} = \frac{(\alpha r_0)^2}{4\pi^2}\frac{139}{90^2}\omega^6
\left(3+\frac{\gamma^2(\cos\theta_3-\beta)^2}
{\gamma^2(\cos\theta_3-\beta)^2+\sin^2\theta_3}\right)^2 \times \\ \nnb
\left(1+\frac{160}{139}\frac{\omega^2\sin^2{\theta_3}}{4\gamma^2(\cos{\theta_3}-\beta)^2+3\sin^2{\theta_3}}\right) \times \\ \nnb 
\sqrt{\frac{\sin^2{\theta_3}}{\gamma^2(\cos{\theta_3}-\beta)^2+\sin^2{\theta_3}}}
\frac{\gamma(1-\beta\cos\theta_3)}
{\gamma^2(\cos\theta_3-\beta)^2+\sin^2\theta_3} d\theta_3 d\phi_3
\eeqa
with $\alpha=1/137$ and $r_0 = 2.8\times10^{-13}$~cm.
This corresponds to the unpolarized case that is comparable to
the s-channel scalar-pNGB exchange with $S=1111$ discussed here, 
because the $S=1111$ cross section eventually coincides with that of
the unpolarized photon--photon scattering.
Since the solid angle and the effect of the induction laser field
are common to both the pNGB exchange and the QED process, we have only
to plug the partially integrated QED cross section into Eq.~(\ref{eq_Yi})
in order to get the QED-based scattering yield.
The boundary line of the gray translucent area in Fig.\ref{Fig3} shows $m$--$g/M$, 
at which $\tilde{\sigma}_{ngb} = \tilde{\sigma}_{qed}$ is satisfied.
This equi-cross-section line intersects with the constraint from
a natural requirement that the life time of a pNGB is equal to the age of 
the universe~\cite{strings} at around 
$(m, g/M) = (2\mbox{~keV}, 10^{-11} \mbox{~GeV}^{-1})$.
Although the QED cross section exceeds the pNGB cross section consistent with
a natural dark matter candidate (dash-dotted line) at m=7~keV, 
the search window of $m<2$~keV is still wide-open 
with relatively low QED background levels.

\begin{table}
\begin{center}
\begin{tabular}{lc}
\hline
Lorentz factor to boost CMS energies&  $\gamma = 1.5$ \\
Scattering angle in CMS & $\theta = \pi/4$ \\
Incident angle of induction beam & $\theta_4 = 1.65$ rad \\
Scattering angle of signal photons & $\theta_3 = 0.31$ rad \\
Common f-number of creation beams & $F=100$ \\
Induction beam f-number & $F_4=10$ \\
Common duration time for creation beams & $\tau = {Z_R}_2/c$ \\
Integration time from longest Rayleigh length & $\tau_L = {Z_R}_2/c$ \\
Common energy uncertainty of creation beams & $R=5$\%\\
Collision repetition rate & $f=1$~Hz \\
Data accumulation time & $T=10^6$~s\\
Total number of signal photons & Y = 100 \\
Detector efficiency to signal photons & $\epsilon = 100$\% \\
\hline
Creation beam energy in CMS & $\omega = 3.50$ keV \\
Higher creation beam energy in ACS & $\omega_1 = 9.16$ keV \\
Lower creation beam energy in ACS & $\omega_2 = 1.34$ keV \\
Induction beam energy in ACS & $\omega_4 = 2.48$ keV \\
Signal photon energy in ACS & $\omega_3 = 8.02$ keV \\
\hline
\end{tabular}
\end{center}
\caption{
Three-beam parameters in CMS and ACS (laboratory frame).
The second set of rows gives example values when tuned at $m=2\omega=7$~keV.}
\label{Tab1}
\end{table}

\section{Light sources}
There are already several short-pulsed coherent/incoherent
light sources that cover the range 0.1~eV to 10~keV. 
In the 1--10~keV range, free-electron lasers are already available.
For instance, an X-ray free-electron laser (XFEL), SACLA~\cite{SACLA}, 
can provide $N_k \sim 10^{11}$ at $f=60$~Hz in
that energy range with an undulator length of 90~m.
Although introducing three long XFEL lines would likely be difficult
from a practical point of view, incoherent photon--photon collisions
for only the creation part combined with an XFEL for the stimulation
part would be the least time-consuming approach. 
This is because incoherent light sources can be attainable using relatively 
compact all-optical laser systems~\cite{PTEP-GG}.
In addition, such a long-scale undulator might be drastically 
shortened by future developments in compact coherent X-ray sources,
such as a graphene-based undulator~\cite{Graphene}.
In the 10~eV--1~keV range, the generation of higher harmonics by shooting
high-intensity laser pulses into material targets
could be useful~\cite{HHG}. 
In the 0.1--10~eV range, variable-wavelength
lasers based on optical parametric amplification are available
as commercial products.
State-of-the-art laser facilities such as the 
Extreme Light Infrastructure~\cite{ELI} can reach
numbers of optical coherent photons per pulse beyond $N_k \sim 10^{20}$ 
with $\tau_L \sim 10$~fs.
These facilities can generate multi-wavelength laser fields up to the keV range
with high intensity by combining several of the methods mentioned above.
\section{Conclusion}
We formulated a stimulated photon--photon scattering
process via an s-channel pNGB exchange in an asymmetric head-on collision 
system that would be applicable to the mass range of 0.1~eV to 10~keV.
Above $m = 2$~ keV, we found that the
QED photon--photon scattering cross section dominates that of a pNGB exchange
whose lifetime is consistent with the age of the universe.
Especially in the 0.1-100~eV range, the domain with
$g/M < 10^{-12}$~GeV${}^{-1}$ has not been tested against any observations 
to date. Therefore, the proposed method could provide a wide search window onto
an unexplored valley in the sensitivity curve with relatively suppressed
QED background levels.
The sensitivity could eventually reach the domain beyond the GUT scale
$M \sim 10^{16}$~GeV, hence, our proposal could provide opportunities to test 
string-theory-based models~\cite{KK}.
 
It would be valuable for future experiments to tune the beam energy and
intensity so that the expected sensitivity could reach 
the equi-cross-section line. Confirmation of the QED process is firstly
important to guarantee that the experiment is indeed performing properly.
Once the QED scattering has been confirmed, it would be interesting to test whether
interference exists between the pNGB exchange and the QED process,
for example, by looking at the angular distribution 
of the signal photons by changing the incident angle of the induction laser 
beam and also the dependence on the combination of linear polarization states.
By repeating this test over four orders of magnitude in the CMS energies,
if we were to see a statistically significant deviation from the QED prediction
in a particular mass range, we could reduce the laser intensity at which the QED effect 
becomes insignificant. If we still see the scattering phenomenon, 
we would be able to claim that something dark is exchanged in 
the photon--photon scattering process.

\begin{acknowledgments}
We thank Y. Fujii for helpful discussions.
K. Homma acknowledges the support of the Collaborative Research
Program of the Institute for Chemical Research,
Kyoto University (Grants Nos.\, 2016--68)
and the Grants-in-Aid for Scientific Research
Nos.\, 15K13487 and 16H01100 from MEXT of Japan.
\end{acknowledgments}


\end{document}